# Continuous-wave amplitude control via the interference phenomenon in acoustic structures

Bingyi Liu[1,*,†], Shanshan Liu[2,*], Liulin Li[1], Chuanxing Bi[3], Kai Guo[1], Yong Li[2,‡], Zhongyi Guo[1,§]

[1]*School of Computer Science and Information Engineering, Hefei University of Technology, Hefei 230009, China*
[2]*Institute of Acoustics, School of Physics Science and Engineering, Tongji University, Shanghai 200092, China*
[3]*Institute of Sound and Vibration Research, Hefei University of Technology, Hefei 230009, China*

**Abstract**

We propose a strategy to continuously tune the amplitude of acoustic waves based on the interference among two mode-conversion paths in passive acoustic structures. The interference phenomenon is attributed to two conjugate acoustic geometric phases obtained with two mode-conversion processes in hybrid-type geometric-phase meta-atom (HGPM) pair. Notably, 100% modulation depth of the wave amplitude is achievable by simply varying the local orientation angle of meta-atom. We utilize the acoustic structure made of two cylindrical resonators to construct deep-subwavelength secondary source with designated initial phase delay, and HGPM supporting desired mode-conversion functionality is accordingly fabricated with four secondary sources. Both theory and experiment consistently verify the continuous amplitude modulation function of HGPM pair, which showcases a general scheme for reconfigurable amplitude-type acoustic meta-devices, i.e., those that require grayscale amplitude modulation for acoustic field engineering.

---

[*] These two authors contributed equally to this work.
[†] E-mail: bingyiliu@hfut.edu.cn
[‡] E-mail: yongli@tongji.edu.cn
[§] E-mail: guozhongyi@hfut.edu.cn

## I. INTRODUCTION

Acoustic metamaterials and metasurfaces open a new avenue to control the acoustic wave with high flexibility by leveraging the artificially designed meta-atoms of subwavelength footprint. Such acoustic meta-atoms could comprehensively manipulate the amplitude, phase and even the frequency of the incident acoustic waves with unprecedented precision via a low-cost and all-passive planar configuration [1,2]. One preferred scheme for the acoustic meta-atoms design relies on the effective medium theory to tune the equivalent acoustic parameters, e.g., the coiling channel [3-7] and Helmholtz resonators [8-11]. However, the intrinsic weaknesses of resonant-type acoustic meta-atoms exactly originate from their resonance nature, which inevitably involves the issues like the coupled amplitude-phase modulation, nonexplicit mapping between the tuned geometry parameters and target phase/amplitude modulations, and limited operating bandwidth, etc. In addition, the low reconfigurability of the state-of-art acoustic meta-atoms hinders their applications in industry and real-life scenarios, which usually require the multi-functionality and dynamic field engineering capability of meta-devices. So far, solution to the tunable acoustic meta-devices has been widely discussed and mainly focuses on tuning the cavity size of the resonant structures, which still fails to circumvent the unexplicit mapping between the geometry parameters and phase/amplitude modulation [12-16]. Other solutions like cascading multiple metasurfaces [17-21] or using non-Hermitian systems [22-24] could only support limited functionalities and loss the generality in constructing pixel-level reconfigurable devices.

Recently, the concept of geometric-phase metasurfaces has been successfully extended from optics to acoustics [25]. By utilizing the mode conversion between two acoustic vortices of different topological charge (TC), a linear mapping between the transmitted/reflected phase and the variance of local orientation angle of acoustic structure is obtained, which is named as acoustic geometric phase (GP). Acoustic GP is analytically given as $\exp\left[j\left(q^{in}-q^{out}\right)\theta\right]$, $q^{in}$ and $q^{out}$ refer to the TC of input and output vortex, $\theta$ is the variance of the orientation angle of the acoustic structure performing the vortex mode conversion, such as phase-gradient metasurfaces [26] and nonlocal acoustic meta-gratings [27]. Notably, the expression of acoustic GP is highly consistent with the optical GP of the form $\exp\left[j\left(\sigma^{in}-\sigma^{out}\right)\varphi\right]$, where $\sigma^{in}$ and $\sigma^{out}$ refer to the spin of input and output circularly polarized light, $\varphi$ is the variance of the orientation angle of anisotropy nano-antenna performing the polarization conversion. Such similarity between the expression of acoustic GP and optical GP is attributed to the SU(2) rotation in an orthogonal coordinate system which eventually gives the helicity-dependent phase modulation. Generally, the helicity can be derived from the spatial inhomogeneity of field (e.g., spiral phase front of acoustic or optical vortex beams) or circular polarizations (i.e., spin of light). Recently, the conversion process among acoustic vortex beams is mapped to and represented by a closed loop on Bloch sphere, which is an analogy to the Poincaré sphere in optics [28]. Based on this picture, people could straightforward understand the appearance of acoustic GP and calculate it with the solid angle formed by a closed loop on the Bloch sphere, and the closed loop represents the evolution of the acoustic beam as it passes through the acoustic structures.

Beside controlling the phase of incident acoustic wave, it is also necessary to manipulate the amplitude of acoustic wave, especially for the application like complex-amplitude acoustic hologram supporting the high-precision field reconstruction, which potentially facilitates better performance on the particle manipulations or ultrasonic therapy. Although acoustic GP enables the acoustic meta-atom with linearly-encoded and tunable phase modulation capability, this strategy fails to tune the amplitude of acoustic waves. So far, the reported acoustic meta-atoms developed for amplitude rectification usually introduce the external loss or rely on the open waveguide of varied aperture size [29,30]. However, these solutions are not easy to operate in a reconfigurable way. Therefore, it is highly desired to find an efficient technique to continuously control the amplitude of acoustic wave via a simple and robust operation that is governed by a universal physical mechanism.

In this work, we provide a solution to continuously control the amplitude of the acoustic wave by taking advantage of the interference of two mode-conversion paths. Different from our previously reported geometric-phase meta-tom that only supports one-to-one mode conversion between two vortices of different TCs, here we integrate two mode-conversion processes within one acoustic meta-atom, i.e., the Hybrid-type Geometric-Phase Meta-atom (HGPM), and continuously tune the amplitude of the transmitted acoustic wave via the interference empowered by two opposite acoustic GPs. Results of the simulations and experiment measurements agree well with our theory, where a continuous modulation of the amplitude of the transmitted acoustic wave with a 100% modulation depth is achieved. Moreover, such mode-interference strategy has no requirement on the minimal pixel size, because the mode-conversion process could be readily realized via near-field coupling mechanism. As a result, acoustic meta-atom of both deep-wavelength geometry and reconfigurable grayscale amplitude modulation functionality is available. Our work reveals another principle to tune the amplitude of acoustic wave, which is robust and reconfigurable, paving the way to construct programmable acoustic amplitude mask for versatile acoustic field engineering scenarios.

## II. RESULTS AND DISCUSSION

### A. Principle of HGPM pair

Figure 1(a) shows the general schematic of our HGPM pair. Here, an individual HGPM is supposed to generate two superimposed acoustic vortex beams with opposite TCs under plane wave illumination, and two vortex modes are denoted by $|\pm q\rangle$, $\pm q$ is the value of TC. Therefore, this meta-atom exactly operates like two phase gradient metasurfaces (PGMs) whose intrinsic topological charges (ITCs) are $\pm q$. Due to the time-reversal symmetry, above meta-atom could also convert the vortex modes $|\pm q\rangle$ to plane wave $|0\rangle$. Then, two mode transformation process $|0\rangle \to |q\rangle \to |0\rangle$ and $|0\rangle \to |-q\rangle \to |0\rangle$ can be constructed by cascading these two identical meta-atoms with proper gap distance $g$. Apparently, these mode-conversion processes are independent to each other and could generate specific acoustic GP modulation as we twist it to tune the relative orientation angle $\theta$ between the two cascaded meta-atoms, see the yellow arrow of Figure 1(a). As a result, the overall transmitted fields can be analytically expressed as:

$$p_T = p_+ + p_- = t_0 p_0 \left(e^{jq\theta} + e^{-jq\theta}\right) e^{jk_z z} = 2 t_0 p_0 \cos(q\theta) e^{jk_z z} \tag{1}$$

here, $p_T$ is the total transmitted field, $p_\pm$ refer to the mode conversion paths involving the mode $|\pm q\rangle$, and the corresponding GPs are $\exp(\pm jq\theta)$, $t_0$ is amplitude of the the transmission coefficient of the cascaded meta-atom system, $p_0$ is the pressure amplitude of incident acoustic wave, $k_z$ is the propagating constant of plane wave transmitted along z direction, which is equal to the operating wavenumber. Therefore, the amplitude of the acoustic wave transmitting through the cascaded meta-atom system is analytically given as $2 t_0 p_0 |\cos(q\theta)|$, which indicates a 100% modulation depth of the amplitude is achievable.

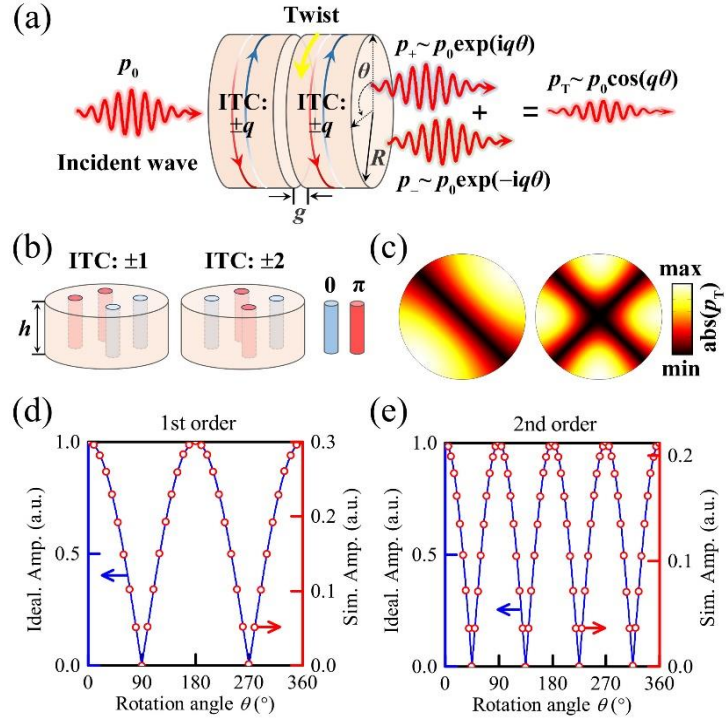

FIG. 1. Principle of the work. (a) Schematic of mode interference when twist the HGPM pair. (b) Configuration of an ideal HGPM supporting the vortex generation of $|0\rangle \to |\pm 1\rangle$ and $|0\rangle \to |\pm 2\rangle$.

(c) The amplitude of acoustic pressure field at the transmitted side of an individual HGPM. Amplitude modulation of the HGPM pair supporting the mode-conversion processes (d) $|0\rangle \rightarrow |\pm 1\rangle \rightarrow |0\rangle$ (1st-order) and (e) $|0\rangle \rightarrow |\pm 2\rangle \rightarrow |0\rangle$ (2nd-order).

Generally, acoustic vortex mode-conversion process is governed by the sound vortex diffraction (SVD) in cylindrical PGM [31,32], which could be regarded as the counterpart of the diffraction law of the acoustic wave incident on a planar PGM in Cartesian coordinate, and the normalized phase gradient of planar PGM is analogous to the ITC of a cylindrical PGM. Similarly, the hybrid-type meta-atom generating two acoustic vortex beams of opposite TCs can be obtained by analogy with binary phase-type meta-grating in Cartesian coordinate, which is a simplified version of the planar PGM that could support ±1 order diffraction. In this work, the HGPM, i.e., a binary phase-type grating in cylindrical coordinate, is achieved with four acoustic secondary sources of initial phase delay 0 or π. Figure 1(b) shows two ideal configurations of the HGPM supporting the generation of $|\pm 1\rangle$ or $|\pm 2\rangle$ vortex pair, of which the four acoustic secondary sources are correspondingly arranged by the order of "0-0-π-π" and "0-π-0-π". Here, we simplify the acoustic secondary sources of HGPM as deep-subwavelength cylindrical waveguides of radius $r$ and height $h$, and fill it with impedance-matched and homogeneous acoustic medium whose acoustic refractive index is $n_0 = 1$ and $n_\pi = 1 + \lambda/2h$, here λ is the operating wavelength. Therefore, when acoustic wave transmits through above waveguides, two acoustic secondary sources with phase delay difference of π is obtained, i.e., the "0" and "π" unit cylinder in color of light blue and red. However, beside the SVD based on the propagating vortex modes, the mode-conversion process can also be realized by a near-field coupling mechanism [26], and the transmission coefficient $t_0$ given in Eq. (1) should be a function of the coupling distance $g$. Therefore, we can scale down the pixel size of meta-atom to a much more subwavelength footprint and the great compactness of the HGPM pair is guaranteed for further field engineering applications.

As a theoretical demonstration, Figure 1(c) shows the amplitude profile of the near-field transmitted pressure field by illuminating an individual HGPM of radius $R =$ 0.25 λ and height $h$ = 0.5 λ, and the radius of secondary source unit is $r$ = 0.25 $R$. Here, the transmitted acoustic wave is an evanescent field, and its lateral field distribution possesses the key merit of the field of superimposed vortex pair. Similar spatial structure features can also be observed in the $TEM_{10}$ and $TEM_{11}$ Hermite-Gaussian modes, which are recognized the sum of two Laguerre-Gaussian modes of opposite TCs [33]. When we cascaded two identical meta-atoms into a pair as Figure 1(a) shows, the transmitted amplitude modulation is illustrated in Figure 1(d, e), which is based on the mode-conversion processes $|0\rangle \rightarrow |\pm 1\rangle \rightarrow |0\rangle$ and $|0\rangle \rightarrow |\pm 2\rangle \rightarrow |0\rangle$, respectively. Here, the red circle plot is the data obtained with the full-wave simulation of ideal HGPM pair, and the blue rigid line refers to the theoretical predictions given by Eq. (1) (taking $q = 1$ and 2). It is apparent that the continuous amplitude control of 100% modulation depth is achieved by simply varying the local orientation $\theta$ angle of top meta-atom. In our simulations, the operating wavelength is set as 10 cm, and the gap distance $g$ for 1st-order and 2nd-order HGPM pair is 2 cm and 1 cm, respectively. Notably, the strategy to scale down the pixel size to deep subwavelength also prevents the existence of unwanted high-order diffractions, which utilizes the waveguide itself as a low-pass filter.

### B. Structure design

Next, we demonstrate our HGPM with realistic acoustic structures. Figure 2(a) shows the cross section of the fundamental unit that functions as an acoustic secondary source, and such configuration has previously been investigated as effective point sources for vortex generation [34]. The unit of height $H$ is composed by two cascaded cavities of diameter $D_{cav}$, and three straight pipes of diameter $D_p$, the height of top and bottom pipe is $h_0$, and the wall thickness is $th$. In our design, the diameter of cavities and connecting pipes, the height of top and bottom pipe, and the total height of unit are initially set as 20 mm, 6 mm, 6 mm, and 92 mm, respectively. The phase of the spherical wave emitted from the secondary source structure can be tuned by varying the height of both top cavity ($h_1$) and bottom cavity ($h_2$). The optimization of the height of two cavities is conducted by integrating the genetic algorithm with the full-wave simulations, and the optimized height $(h_1, h_2)$ of the secondary source corresponding to the 0 and π phase delay are (26.8, 25.8) mm and (10.6, 14.4) mm, respectively. Figure 2(b) shows a 3D schematic of 1st-order HGPM, here the radius $R$ is set as 53.35 mm to match the realistic impedance tube whose inner radius is 50 mm, and the lateral shift distance $d_s$ of four point-source units is accordingly optimized as 19.4 mm. In our theoretical demonstration, we aim to determine the optimized operating frequency and finely tune the value of $d_s$ based on the transmission minimum of single HGPM, of which the direct transmission is expected to be efficiently suppressed and only near-field coupling contributes to the energy transfer among the HGPM pair. Based on this, the optimized central operating frequency of 1st and 2nd order HGPM is 1210 Hz and 1230 Hz, respectively.

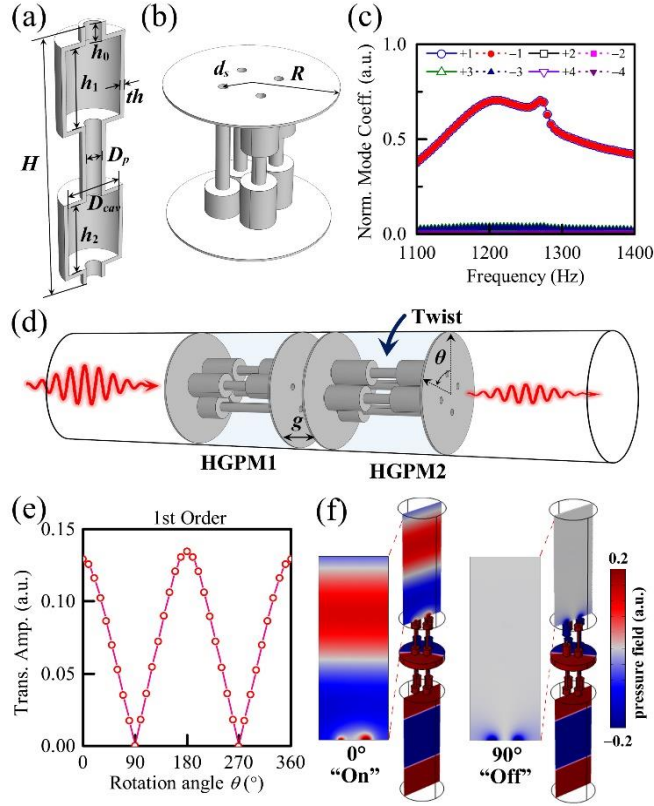

FIG. 2. Realization of HGPM pair. (a) Schematic of fundamental secondary source unit and (b) 1st-order HGPM. (c) Coefficients of the vortex modes with TC ranging from ±1 to ±4 over the frequency range from 1100 Hz to 1400 Hz. (d) Acoustic waveguide with HGPM pair. (e) Amplitude modulation of HGPM pair by varying the local orientation angle $\theta$, where the coupling distance $g$

is 1.5 cm. (e) Acoustic pressure field of HGPM pair whose twist angle is 0° and 90°.

As a straightforward demonstration of the vortex generation performance of designed HGPM, we leverage the orthogonality property of vortex beams of different TCs and conduct mode expansion on the near field obtained at the transmitted end of the structure. Figure 2(c) shows the normalized coefficient of the evanescent vortex modes whose TC range from ±1 to ±4 over the frequency range from 1100 Hz to 1400 Hz. It is obvious that the vortex modes of TCs ±1 (hollow and solid circles) are the dominant components while other vortex modes can be neglected. Figure 2(d) is the schematic of HGPM pair within a fitted waveguide, here the gap $g$ between meta-atoms is 1.5 cm, and the top HGPM can be rotated by angle of $\theta$. Figure 2(e) shows the transmitted amplitude of 1$^{st}$-order HGPM pair as we varying $\theta$ from 0° to 360°. It is apparent that the transmitted amplitude continuously oscillates between a maximum and 0 in the form of function $|\cos(\theta)|$, of which the modulation depth is 100%. As we decrease the coupling distance $g$, the maximum transmitted amplitude can be increased as well, however, the stronger near-field coupling would distort the line shape away from the theory predictions. Figure 2(f) shows the acoustic pressure field of two classic states, i.e., "On" and "Off" state obtained at the twist angle of 0° and 90°.

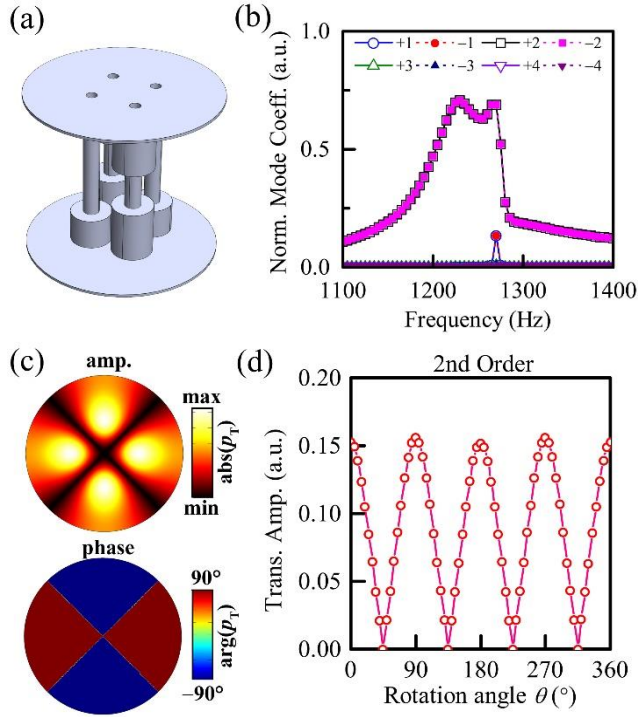

FIG. 3. High-order HGPM pair. (a) Configuration of 2$^{nd}$-order HGPM. (b) Coefficient of the vortex eigenmodes with TC ranging from ±1 to ±4 over the frequency range from 1100 Hz to 1400 Hz. (c) The amplitude profile at the cut-plane above surface of an individual HGPM under the plane wave incidence at 1230 Hz. (d) Amplitude modulation of meta-atom pair by varying the local orientation angle of top HGPM, where the coupling distance $g$ is 1 cm.

We further investigate the HGPM pair relying on higher-order mode-conversion processes. Figure 3(a) is the configuration of 2$^{nd}$-order HGPM, whose building secondary sources are the same as that given in Figure 2(a) but rearranged in the order of "0-π-0-π". In this scenario, the mode-conversion processes $|0\rangle \to |\pm 2\rangle \to |0\rangle$ are utilized. Figure 3(b) shows the normalized coefficient of the evanescent vortex modes generated by an individual meta-atom over the frequency range from 1100 Hz to 1400 Hz, where we could observe strong vortex modes of TCs ±2 (hollow and solid cubes). Figure 3(c) shows the amplitude and phase profile of the near-field acoustic pressure field obtained at the transmitted side of an individual 2$^{nd}$-order HGPM, where we could observe typical spatial features of the interference of two vortex modes, especially the binary phase distribution. Moreover, Figure 3(d) shows the transmitted amplitude as the local orientation angle $\theta$ of top HGPM is increased from 0° to 360°, whose oscillation period is 45°. Here, we only test the coupling distance $g$ of 1 cm, and smaller $g$ generally could allow for higher energy transmission but may distort the line shape from ideal cosine function.

### C. Experimental demonstration

We fabricate the sample with state-of-art 3D printing technique where the wall thickness $th$ of the structure is set as 2.5 mm to make the structure to be sound hard enough. In the experiment, a Brüel & Kjær type-4206T impedance tube with a diameter of 100 mm is used to measure the transmission coefficients of the samples. The loudspeaker generates a plane incident wave, and the local pressure fields captured by four 1/4-inch condenser microphones (Brüel & Kjær type-4187) are utilized to retrieve the transmission coefficients. The picture of real sample is shown in Figure 4(a), and the experiment setup is schematically given in Figure 4(b). In this work, we also fabricate the adaptor waveguide with 3D printing to better fit the impedance tube and HGPM pair. Notably, in our experiment, we found that the overall transmission of HGPM pair is low when compared to theoretical calculation. Such results are understood as the leakage of acoustic energy at the junctions among different components (e.g., HGPM structure, waveguide adaptor, etc.) and the dissipation caused by the thermal acoustic effects in narrow connecting regions. In this case, instead of finding the transmission dip of single HGPM, we pick up the optimized working frequency at the transmission peak of HGPM pair. Figure 4(c,d) shows the measured transmitted amplitude of $1^{st}$-order and $2^{nd}$-order HGPM pair, whose gap size are 1 cm and 0.75 cm, and the optimized operating frequency are measured as 1238 Hz and 1269 Hz, which are close to the peak frequencies given by the simulation considering the energy leakage. In our measurement, we vary the orientation angle $\theta$ by step of 10°, while additional four angle value (45°, 135°, 225°, 315°) are added into the measurement of $2^{nd}$-order HGPM pair. It is apparent that the amplitude of transmitted acoustic wave indeed oscillates from its maximum value (>0.1) to 0, whose modulation depth is nearly 100%. Therefore, continuous amplitude control of acoustic waves is available with our reconfigurable HGPM pair.

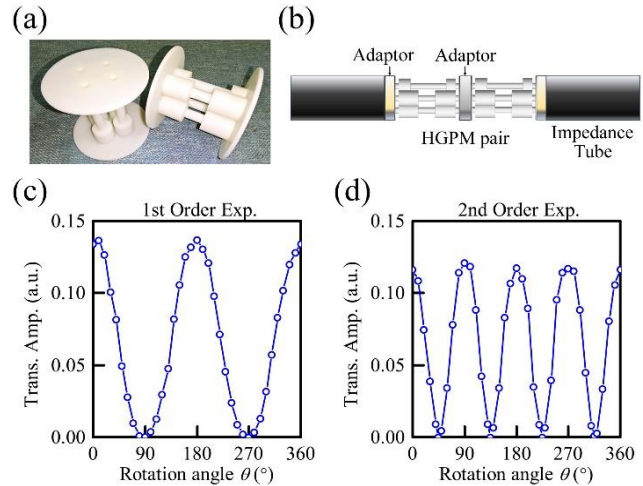

FIG. 4. Experimental verification. (a) Fabricated samples of $1^{st}$ order HGPM. (b) Experiment setup. Measured transmitted amplitude of (c) $1^{st}$-order and (d) $2^{nd}$-order HGPM pair at the optimized operating frequency.

### D. Discussions

Our HGPM pair made of the resonant units is a proof of principle of the geometric-phase induced interference phenomenon in artificial acoustic system, where a nearly 100% modulation depth of the wave amplitude can be easily achieved with unprecedented precision and extreme simplicity, i.e., by only rotating the HGPM structure. In theory, such amplitude modulation is free of additional phase modulation or involving any strict geometry parameters as that required by open waveguides [30]. It should be noted that, our HGPM pair shows the potential to operate in a broadband manner, which can be deduced from the mode expansion results retrieved from the near field of single HGPM. Considering its potential in constructing programmable acoustic devices, small HGPM diameter is needed to facilitate mechanical transmission structure connecting the stepper motor to drive it. Therefore, in the future, it is necessary to develop other design strategies that could realize compact assemble of dense amplitude-tunable acoustic pixels while maintain a higher transmission maximum of the acoustic energy.

## III. CONCLUSIONS

In this work, we propose a robust and feasible solution for acoustic wave amplitude control, which merely relies on the geometric nature of HGPM pair. Based on the concept of interference among the mode-conversion paths, the two conjugate geometric phases associated with the mode-conversion processes give a cosine function modulation on the transmitted acoustic waves. Both theoretical calculations and experimental measurements verify such rotation-angle-dependent amplitude modulations of 100% modulation depth. Our HGPM pair opens a new avenue to control the amplitude degree of freedom of waves, and it can be potentially applied in programmable acoustic field engineering that requires complex amplitude modulations of acoustic waves.

**ACKNOWLEDGEMENTS**

B.L. acknowledged the financial support of the National Nature Science Foundation of China (Grant No. 12104044). G.Z. acknowledged the financial support of the National Nature Science Foundation of China (Grant No. 61775050). B.L. acknowledged the help offered by Dr. Yuhong Na (M) from Anhui University in polishing the language of the paper. Y.L. acknowledges the Xiaomi Young Talents Program.